\documentclass[a4paper,fleqn]{cas-sc}

\usepackage{amssymb}
\usepackage{amsmath}
\usepackage{amsthm}

\usepackage[T1]{fontenc}
\usepackage{algorithm}
\usepackage{algorithmic}
\usepackage{array}
\usepackage[caption=false,font=normalsize,labelfont=sf,textfont=sf]{subfig}
\usepackage{textcomp}
\usepackage{url}
\usepackage{verbatim}
\usepackage{graphicx}
\usepackage{booktabs}
\usepackage{multirow}
\usepackage{hyperref}
\usepackage{xcolor}
\usepackage{enumitem}
\usepackage{placeins}
\usepackage{microtype}
\usepackage{xurl}

\usepackage[numbers]{natbib}

\begin{document}
\let\WriteBookmarks\relax
\def\floatpagepagefraction{.8}
\def\textpagefraction{.1}
\renewcommand{\topfraction}{0.9}
\renewcommand{\bottomfraction}{0.8}
\renewcommand{\floatpagefraction}{0.8}

\shorttitle{EvoMarket}
\shortauthors{Zhong et~al.}

\title[mode = title]{EvoMarket: A High-Fidelity and Scalable Financial Market Simulator}

\author[aff1,aff2]{Muyao Zhong}[
  orcid=0009-0008-9878-4435
]

\author[aff2,aff3]{Zhenhua Yang}

\author[aff4]{Yuxiang Liu}

\author[aff2]{Ke Tang}[
  orcid=0000-0002-6236-2002
]

\author[aff4,aff2]{Peng Yang}[
  orcid=0000-0001-5333-6155
]
\cormark[1]
\ead{yangp@sustech.edu.cn}
\cortext[cor1]{Corresponding author}

\affiliation[aff1]{organization={Department of Electronic Information, Harbin Institute of Technology},
            city={Harbin},
            postcode={150001},
            country={China}}

\affiliation[aff2]{organization={Guangdong Provincial Key Laboratory of Brain-Inspired Intelligent Computation, Department of Computer Science and Engineering, Southern University of Science and Technology},
            city={Shenzhen},
            postcode={518055},
            country={China}}

\affiliation[aff3]{organization={Zhongguancun Academy},
            city={Beijing},
            postcode={100094},
            country={China}}

\affiliation[aff4]{organization={Department of Statistics and Data Science, Southern University of Science and Technology},
            city={Shenzhen},
            postcode={518055},
            country={China}}

\begin{abstract}
High-fidelity, scalable market simulation is a key instrument for mechanism evaluation, stress testing, and counterfactual policy analysis. Yet existing simulators rarely achieve \emph{mechanism fidelity} beyond single-asset intraday settings, \emph{microstructure fidelity} against historical limit order books (LOB), and \emph{computational tractability} at market scale in a single system. This paper presents \textit{EvoMarket}, a discrete-event, multi-agent financial market simulator designed for intervention-oriented experiments in multi-asset and cross-day environments. EvoMarket couples a high-throughput execution core (optimized LOB data structures, hierarchical scheduling under propagation delays, and asynchronous per-asset matching) with explicit institutional mechanisms (market calendars, opening call auctions, price limits, and T+1 settlement). To avoid expensive black-box calibration, EvoMarket introduces an Oracle-guided in-run self-calibration mechanism that interprets microstructure discrepancy as missing order flow and synthesizes corrective orders at recording checkpoints. Experiments on China A-share order-flow and LOB data show close replay alignment over five trading days, fidelity gains from budgeted in-run calibration across depth levels, broad agent order-space coverage, and scalable performance under increasing input order rates and market breadth. We further demonstrate cross-asset linkage and event-study style intervention evaluation that produces structured dependence and interpretable event-time responses.
\end{abstract}

% \begin{highlights}
% \item Multi-asset, cross-day simulation with explicit institutional mechanisms.
% \item In-run self-calibration converts LOB gaps into corrective orders (no outer loop).
% \item Reaches tick-level price accuracy within one run (12 s for a 1 h window).
% \item Single-core processes 50k orders/s in 1.77 s; supports linkage and interventions.
% \end{highlights}

%% Keywords

\maketitle

\section{Introduction}\label{sec:intro}

Financial markets support economic growth, financial security, and public welfare, making market stability a top-tier concern\cite{zheng2024new,hussain2011probabilistic}. At the same time, markets operate as large-scale socio-technical systems in which heterogeneous participants, including institutional investors, retail traders, market makers, and regulators\cite{daah2025simulation}, interact under explicit trading rules\cite{hasbrouck2007empirical}. These interactions generate nonlinear feedback and emergent dynamics such as bubbles\cite{denatale2026multi}, contagion\cite{allen2000contagion}, liquidity dry-ups\cite{brunner2008liquidity}, and volatility clustering\cite{zhang2010modeling,li2023analyzing}. Such dynamics can amplify shocks and propagate stress across institutions and assets, increasing the likelihood of systemic instability\cite{haldane2011systemic}.

Rigorous study of such mechanisms is challenging because many questions of interest are inherently \emph{counterfactual}\cite{imbens2024causal}. Let $a\in\mathcal{A}$ denote an intervention that specifies a market regime, such as a trading rule set, an agent behavior profile, or a regulatory action with a given timing and intensity. For each $a$, the market system induces an outcome trajectory $Y(a)$, while real-world records reveal only the realized trajectory $Y(a_0)$ under the historically deployed regime $a_0$. Counterfactual analysis seeks to infer $Y(a)$ for $a\neq a_0$ or to quantify an effect of changing regimes, for example $\Delta(a,a_0)=\Psi\big(Y(a),Y(a_0)\big)$, where $\Psi(\cdot)$ is an application-dependent comparison functional (e.g., an event-study contrast or a difference-in-differences style summary over event-time trajectories)\cite{angrist2014mastering}. Empirical analysis based on $Y(a_0)$ (e.g., backtesting\cite{campbell2005backtest}) can be informative about associations, but it provides limited leverage for isolating mechanisms under alternative rules or behaviors\cite{luo2025concentrated}. Moreover, endogeneity and confounding shocks, together with the infeasibility of controlled repeated interventions, make causal attribution difficult using observational data alone\cite{hasbrouck2007empirical}.

These challenges motivate the development of simulation instruments that serve as testbeds for research\cite{xue2022computational}. 
In this sense, the simulator should make it feasible to generate and compare trajectories $Y(a)$ in a range of interventions $a$, with controlled stochasticity and rich instrumentation. 
Concretely, we require that the simulator (i) remains \emph{computationally tractable} so that $Y(a)$ can be produced for practical market scope and high event rates, (ii) is \emph{controllable} so that $a$ can encode rule changes and behavioral shifts of interest, (iii) is \emph{observable} by exposing microstructure-level states (i.e., limit order book, LOB\cite{gould2013lob}), and (iv) is \emph{reproducible} by supporting repeatable, comparable runs for fixed configurations and principled comparative evaluation\cite{xue2024computational2}.
Beyond these requirements, practical quantitative workflows also demand multi-asset capability\cite{g2024enhancing}. Many tasks in modern quantitative trading, such as alpha factor mining, universe construction, and portfolio-level execution and hedging, rely on cross-sectional information and cross-asset constraints, and therefore cannot be evaluated meaningfully in isolated single-asset simulators whose outputs are independent by construction.

In practice, existing tools satisfy only a subset of these requirements. Brokerage paper-trading platforms\cite{schwab_paper_trading} and internal matching sandboxes can execute production-like order handling, but they are often proprietary and provide limited microstructure observability and limited support for rule-level interventions. The NASDAQ exchange test environment\cite{nasdaq_ntf_guide} is closer to production protocols, yet access and allowed interventions are restricted, limiting reproducible comparison at market scale. Here we use \emph{market scale} to denote settings with thousands of listed stocks (e.g., the Shenzhen Stock Exchange, or SZSE with more than 2{,}000 stocks) and sustained high event rates (on the order of tens of thousands of orders per second), which collectively require multi-asset, cross-day simulation. Laboratory or online experimental markets\cite{hendershott2022transparency} allow controlled treatments, but they operate at much smaller scale and abstract away institutional details of modern electronic markets. Finally, order replay and historical backtesting\cite{balch2019evaluate,bailey2017probability} evaluate only the realized regime $Y(a_0)$ and therefore cannot generate or compare counterfactual trajectories under alternative rules or behaviors.

Academic market simulators, such as agent-based interactive discrete event simulation (ABIDES)\cite{byrd2019abides} and 
Multi-Agent eXchange Environment (MAXE)\cite{belcak2022maxe}, 
have enabled more controlled market experimentation by providing executable market mechanisms and programmable agent behaviors. 
Nonetheless, it remains challenging to achieve high-fidelity simulation at market scale in practical computational budgets when simultaneously aiming at multi-asset interaction, cross-day mechanisms, market-specific rules, and close alignment to historical LOB statistics\cite{frey2023jax}. Existing systems are typically designed and evaluated with only a subset of these requirements in mind. For example, ABIDES emphasizes the execution of the mechanism and the programmability of the agent in a single-asset intraday setting\cite{byrd2019abides}. MAXE is optimized primarily for throughput and high event rates\cite{belcak2022maxe}, while coverages of broader trading behaviors and end-to-end intervention evaluation are not its main focus of design \cite{frey2023jax}. As a result, two fidelity requirements remain difficult to jointly satisfy.

First, \emph{mechanism fidelity} concerns whether institutional mechanisms and trading rules are faithfully implemented,
including multi-asset interaction and cross-day sessions (e.g., pre-open auctions and intraday breaks), as well as market-specific constraints such as price limits and settlement rules.

Second, \emph{microstructure fidelity} concerns whether the simulator reproduces historical LOB statistics, including spread and depth dynamics, order-flow regularities, and stylized facts\cite{abergelLimitOrderBooks2016,hasbrouck2007empirical,lebaron2006stylized}. Achieving high microstructure fidelity typically requires \emph{calibration}. Given historical observations $\hat{x}_{1:T}$, a simulator $\mathcal{M}$, and parameters $\theta$, a standard formulation is
\begin{equation}
\theta^*=\arg\min_{\theta} \mathcal{D}\big(\Phi(\mathcal{M}(\theta)),\Phi(\hat{x})\big),
\end{equation}
where $\Phi(\cdot)$ summarizes microstructure features from LOB snapshots and order-flow events, and $\mathcal{D}(\cdot,\cdot)$ measures their discrepancy.
Existing black-box approaches (e.g., approximate Bayesian computation (ABC)\cite{goosen2021calibrating} or derivative-free optimization\cite{dyer2024black}) typically require many repeated simulator runs, so wall-clock time grows quickly with the number of evaluations and the per-run horizon. In these methods, each objective evaluation requires simulating the full calibration window, even when the discrepancy $\Phi(\cdot)$ is computed only on a coarse grid such as 1-minute features summaries aggregated from high-frequency order flow. Importantly, reported calibration costs are often measured in \emph{single-asset} settings. For example, based on the runtime reported for sequential Monte Carlo (SMC)-ABC calibration on high-frequency data\cite{goosen2021calibrating}, calibrating a single asset over one trading day with a common low-frequency feature discretization of 1-minute snapshots can require $\sim$9{,}300 minutes (about 155 hours) on a 40-core machine. A broader benchmarking study\cite{platt2020comparison} further reports that fitting larger parameterizations across calibration methods can take from several days to multiple weeks of wall-clock time even for comparably sized datasets. 

This burden becomes prohibitive when calibration targets expand from a single asset to market-wide, multi-asset, cross-day settings with large heterogeneous populations of agents. Taking SZSE as an example (on the order of over 2{,}000 listed stocks), the single-asset, one-day estimate above implies decades of wall-clock time on the same 40-core machine, before accounting for cross-asset coupling and joint calibration targets that typically further increase the cost. Such costs make high-fidelity, market-scale simulation difficult, and cost-driven simplifications can in turn weaken mechanism evaluation and counterfactual analysis.

This tension motivates two core questions.
\begin{enumerate}
  \item \textbf{To what extent can fidelity and scalability be achieved simultaneously in market simulation?}
  \item \textbf{How can such high-fidelity be achieved with affordable computational costs?}
\end{enumerate}

To address these questions, we present \textit{EvoMarket}, a discrete-event, multi-agent financial market simulator. EvoMarket is designed for scalable studies that require multi-asset environments, cross-day operation, and intervention-oriented experimentation.
Besides various mechanism-fidelity improvements, a central design choice is to move calibration from an iterated external outer loop to a simulator-internal evolution process. Rather than repeatedly invoking black-box optimizers that rerun $\mathcal{M}(\theta)$ over the full calibration window for each parameter evaluation, EvoMarket exposes discrepancy signals during execution and uses an \emph{Oracle} to coordinate specific agent groups that heuristically construct corrective actions to reduce data discrepancy at each timestep. This white-box construction exploits simulator structure and substantially avoids expensive outer-loop iterations.

Section~\ref{sec:experiments} shows that EvoMarket makes market-scale experimentation practical by combining mechanism realism, fast microstructure alignment within one run, and near-real-time execution. These results support repeatable experiments that remain fidelity-oriented and computationally affordable as market breadth increases.

We summarize three key features of EvoMarket.
\begin{itemize}[leftmargin=*]
    \item \textbf{Mechanism fidelity for scalable multi-asset and cross-day settings.} EvoMarket executes per-order multi-asset exchanges under explicit session calendars and market-specific constraints, including opening call auctions, price limits, and settlement rules. This mechanism layer enables consistent cross-day replay and rule-level interventions beyond single-asset intraday simulators.
    \item \textbf{Self-evolution to historical microstructure.} The self-calibration mechanism reduces snapshot-level microstructure discrepancy within a single run. In the one-hour calibration benchmark, one EvoMarket run completes in about 12 seconds and reaches root-mean-squared price error smaller than $10^{-1}$ RMB (a few ticks of 0.01 RMB size). Compared with PSO-based outer-loop calibration run for wall-clock comparable to the target window, EvoMarket achieves about $10\times$ lower residual error while using about $1/300$ of the calibration time.
    \item \textbf{High-throughput, scalable execution.} The execution engine sustains market-scale order rates. In the stress test, processing the 50k orders per second workload takes 1.77 seconds wall-clock, approaching real-time execution for the about 30k orders per second regime. Such order rate is observed from SZSE with tens of millions of traders and over 2000 assets. And this efficiency enables simulation with richer heterogeneous agent populations and repeated intervention studies without sacrificing microstructure detail.
\end{itemize}

The paper is organized as follows. Section~\ref{sec:prelim} introduces notation and formal definitions; Section~\ref{sec:related} reviews related work; Section~\ref{sec:system} presents the design of EvoMarket; Section~\ref{sec:experiments} reports fidelity, calibration, scalability, and intervention results. Section~\ref{sec:conclusion} concludes.

\section{Preliminaries}\label{sec:prelim}

This section introduces formal definitions and notation used throughout the paper, including markets, orders, and LOBs. 

\subsection{Market mechanism and exchange}
We consider an electronic market as a mechanism that maps a time-ordered stream of orders into executions and an evolving market state. In modern electronic markets, the central state is the LOB maintained by an exchange under a continuous double auction (CDA) mechanism. CDA matches incoming orders by price-time priority. Buy orders compete to pay higher prices and sell orders compete to accept lower prices, while orders at the same price are prioritized by arrival time\cite{gould2013lob,abergelLimitOrderBooks2016}.

\subsection{Orders}
An order is a message submitted to the exchange. We represent an order as a tuple $o=(\mathrm{id}, j, s, \tau, p, v, t)$, where $\mathrm{id}$ is a unique identifier, $j\in\{1,\ldots,M\}$ is the asset index, $s\in\{\mathrm{buy},\mathrm{sell}\}$ is the side, $\tau\in\{\mathrm{limit},\mathrm{market},\mathrm{cancel}\}$ is the order type, $(p,v)$ denote the price and volume (with $p$ undefined for a market order), and $t$ is the event time at which the exchange receives the order. A limit order specifies a price constraint and adds liquidity if not immediately executable, whereas a market order executes against the best available opposite-side prices until its volume is filled or liquidity is exhausted. A cancel message removes an existing resting limit order identified by $\mathrm{id}$.

\subsection{Agents and portfolios}
Agents may hold and trade multiple assets. Let $\mathbf{h}_{i,t}\in\mathbb{R}^{M}$ denote the share holdings of agent $i$ across $M$ assets at event time $t$, and let $c_{i,t}\in\mathbb{R}$ denote the available cash. A decision at time $t$ may submit a \emph{batch} of orders across assets, denoted by $\mathcal{O}_{i,t}=\{o_{i,t}^{j}\}_{j\in\mathcal{J}_{i,t}}$, where $\mathcal{J}_{i,t}\subseteq\{1,\ldots,M\}$ is the set of assets traded at that decision point. This portfolio-level interface is necessary for modeling cross-asset strategies and constraints and aligns with how real traders manage inventory and risk at scale.

\subsection{LOB snapshots and dynamics}
At any event time $t$, a LOB snapshot records the best $l$ price levels on each side. We represent a snapshot $\mathbf{L}_{t}\in\mathbb{R}^{l\times 4}$ as
\begin{equation}
\mathbf{L}_{t}=
\begin{bmatrix}
b_{1}^{p} & b_{1}^{v} & a_{1}^{p} & a_{1}^{v} \\
b_{2}^{p} & b_{2}^{v} & a_{2}^{p} & a_{2}^{v} \\
\vdots & \vdots & \vdots & \vdots \\
b_{l}^{p} & b_{l}^{v} & a_{l}^{p} & a_{l}^{v}
\end{bmatrix},
\end{equation}
where $b_{i}^{p}$ and $a_{i}^{p}$ denote the bid and ask prices at level $i$, and $b_{i}^{v}$ and $a_{i}^{v}$ denote the corresponding aggregated volumes. The snapshot obeys cross-feature constraints:
\begin{equation*}
b_{1}^{p}>b_{2}^{p}>\cdots>b_{l}^{p}, \quad a_{1}^{p}<a_{2}^{p}<\cdots<a_{l}^{p}, \quad b_{1}^{p}<a_{1}^{p},
\end{equation*}
and all volumes are nonnegative. Over a horizon of $T$ aligned timestamps, the LOB can be viewed as a multivariate time series $\mathbf{S}_{t}=[\mathbf{L}_{t},\mathbf{L}_{t+1},\ldots,\mathbf{L}_{t+T-1}]\in\mathbb{R}^{T\times l\times 4}$.

The evolution of the LOB is induced by the exchange mechanism. Let $\mathcal{O}_{t}^{t+1}$ denote the set of orders, submitted by all agents, arriving between times $t$ and $t+1$, and let $\mathcal{C}$ denote the CDA matching operator. Then the update can be written as
\(\mathbf{L}_{t+1}= \mathcal{C}(\mathbf{L}_{t}, \mathcal{O}_{t}^{t+1}),\)
which makes explicit the dependence between consecutive snapshots and the order stream that drives state transitions.

\section{Related Work}\label{sec:related}

Market simulation has long been studied as a computational instrument for analyzing trading mechanisms and emergent market dynamics under controlled interventions, and it is increasingly framed as a ``computational experiment'' for market mechanism analysis\cite{xue2022computational,lu2022computational2,xue2024computational}. In the notation of Section~\ref{sec:prelim}, the core modeling question is how to construct an executable mechanism $\mathcal{C}$ and a population of heterogeneous decision makers that generate order streams $\{\mathcal{O}_{t}^{t+1}\}$ such that the induced LOB process $\{\mathbf{L}_t\}$ is both faithful to institutional rules and aligned with empirical microstructure statistics.

Early agent-based models (ABMs), including the Santa Fe Institute Artificial Stock Market (SFI-ASM)\cite{ehrentreich2008agent}, demonstrated that heterogeneous and adaptive interacting agents can reproduce stylized facts and provide mechanistic explanations complementary to purely statistical analysis\cite{arthur2018asset,sagwal2025analyzing,lebaron2006stylized}. Modern simulators increasingly adopt discrete-event architectures with explicit exchanges, message latency, and LOB-level message handling to represent electronic trading protocols\cite{byrd2019abides}. ABIDES, for example, provides a programmable agent interface and an exchange-side CDA mechanism, enabling controlled studies of how behavioral assumptions interact with market rules. Throughput-oriented simulators further optimize microstructure kernels to support market-scale experimentation and training, such as GPU-accelerated LOB execution in JAX-LOB\cite{frey2023jax} and the high event-rate discrete-event design in MAXE\cite{belcak2022maxe}. A complementary line of work focuses on reinforcement-learning trading environments, such as the Financial Market Simulation Environment (FMSE)\cite{mascioli2024financial}, which learn agent policies within a fixed simulator interface. These systems collectively illustrate a recurring tension. As experimental scope expands toward multi-asset, cross-day, and high-frequency regimes, both mechanism fidelity and microstructure fidelity become harder to maintain within practical computational budgets.

Mechanism fidelity depends critically on institutional details beyond continuous intraday CDA. Call auctions at the open and close are widely used for price discovery and liquidity aggregation\cite{budish2015high,bogousslavsky2023trades,chen2013modeling}. Stabilization devices such as price limits\cite{aktas2021volatility}, trading pauses\cite{hautsch2019effective}, and circuit breakers\cite{bongaerts2024circuit} further reshape feasible order flow and inventory dynamics, while settlement cycles and related constraints (e.g., T+1/T+2) impose cross-day inventory and execution frictions\cite{madhavan2000market}. Cross-asset interaction is increasingly recognized as a first-class driver of microstructure dynamics through cross-impact and correlated order flow\cite{cont2023cross}. Consequently, simulators intended for policy and systemic-risk studies must represent calendars and session transitions, including the coupling between overnight events and daytime trading\cite{ham2022effects}, rather than treating the market as a stationary single-asset intraday process.

To summarize what is (and is not) covered by representative platforms, Table~\ref{tab:related-checklist} provides a compact checklist of fidelity and scalability dimensions that a market simulator may need to support for intervention-oriented studies.

\newcommand{\cmark}{\ensuremath{\checkmark}}
\newcommand{\xmark}{\ensuremath{\times}}
\newcommand{\pmark}{\ensuremath{\triangle}}

\begin{table*}[pos=t]
  \centering
  \caption{Checklist of representative market simulator platforms aligned with the experimental groups in Section~\ref{sec:experiments}. A checkmark indicates supported, a triangle indicates partially supported or limited scope, and a cross indicates not supported or not a design focus. ``Agents'' indicates support for heterogeneous/custom agents; ``Rules'' refers to institutional mechanisms beyond continuous intraday CDA (e.g., call auctions, price limits, settlement constraints). ``Order scale'' indicates the ability to sustain exchange-scale order rates under practical budgets; ``Agent scale'' indicates the ability to scale to large heterogeneous agent populations.}
  \label{tab:related-checklist}
  \footnotesize
  \setlength{\tabcolsep}{3.5pt}
  \begin{tabular}{lcccccccc}
    \toprule
    & \multicolumn{5}{c}{Mechanism fidelity} & \multicolumn{1}{c}{Microstructure fidelity} & \multicolumn{2}{c}{Scalable efficiency} \\
    \cmidrule(lr){2-6}\cmidrule(lr){7-7}\cmidrule(lr){8-9}
    Work & LOB CDA & Agents & Multi-asset & Cross-day & Rules & Calibration & Order scale & Agent scale \\
    \midrule
    SFI-ASM~\cite{ehrentreich2008agent} & \xmark & \pmark & \xmark & \xmark & \xmark & \xmark & \xmark & \xmark \\
    ABIDES~\cite{byrd2019abides} & \cmark & \cmark & \pmark & \xmark & \xmark & \xmark & \pmark & \pmark \\
    MAXE~\cite{belcak2022maxe} & \cmark & \pmark & \pmark & \xmark & \xmark & \xmark & \cmark & \pmark \\
    JAX-LOB~\cite{frey2023jax} & \cmark & \xmark & \pmark & \xmark & \xmark & \xmark & \cmark & \xmark \\
    FMSE~\cite{mascioli2024financial} & \pmark & \pmark & \xmark & \xmark & \xmark & \xmark & \xmark & \xmark \\
    \midrule
    EvoMarket (this work) & \cmark & \cmark & \cmark & \cmark & \cmark & \cmark & \cmark & \cmark \\
    \bottomrule
  \end{tabular}
\end{table*}

Overall, representative simulators and learning environments typically cover only subsets of these dimensions; EvoMarket is designed to jointly support mechanism fidelity, microstructure fidelity, and scalable execution within a single system.
Institutional mechanisms and multi-asset coupling leave characteristic signatures in order flow and LOB state trajectories, motivating microstructure-level evaluation and calibration. Microstructure fidelity is commonly assessed against historical observables derived from LOB and order-flow statistics\cite{hasbrouck2007empirical,gould2013lob,abergelLimitOrderBooks2016}. Beyond evaluation, learned representations of LOB dynamics (e.g., DeepLOB\cite{zhong2025representation,tian2025graph,li2025simlob}) have been developed for prediction tasks\cite{contreras2018enmx} and can also serve as feature extractors when defining discrepancy maps $\Phi(\cdot)$.

Accordingly, calibration of ABMs and market simulators is often posed as simulation-based inference or optimization. Given a simulator $\mathcal{M}(\theta)$ and observed data $\hat{x}$, one seeks parameters $\theta$ that minimize a discrepancy $\mathcal{D}(\Phi(\mathcal{M}(\theta)),\Phi(\hat{x}))$ (cf.\ Section~\ref{sec:intro}). Representative approaches include approximate Bayesian computation\cite{goosen2021calibrating}, black-box Bayesian inference\cite{dyer2024black}, surrogate-assisted calibration\cite{lamperti2018agent,jiang2026calibrating}, and deep calibration with learned embeddings and density estimators\cite{stillman2023deep}. Recent finance-focused work further explores posterior-assisted online calibration and refined discrepancy objectives for multivariate time series\cite{yang2026posterior,wang2025alleviating}. Benchmarking studies indicate that calibration cost can vary by orders of magnitude and increases sharply with parameter dimension and length of observed data\cite{platt2020comparison}, consistent with broader perspectives from simulation-based inference\cite{cranmer2020frontier}. Besides, when the experimental scope expands to multi-asset and cross-day settings, the main cost driver often becomes repeatedly simulating batches of full trading sessions required by external calibration or optimization routines\cite{fang2026efficient}.

Scalability and throughput are therefore central concerns for high-fidelity simulation at market scale. Parallel discrete-event simulation provides foundational techniques for event scheduling, synchronization, and reproducibility\cite{fujimoto1990parallel,jagtap2012optimization}. GPU-based ABM frameworks demonstrate substantial acceleration when interactions are data-parallel, although exchange-side microstructure often imposes tighter coupling constraints\cite{richmond2023flame}. Recent LOB simulators emphasize hardware acceleration and efficient kernels to support market-scale experimentation and learning\cite{frey2023jax}. These efforts clarify that high-frequency, market-wide simulation is feasible only when the matching critical path, event scheduling, and logging are designed as first-order systems' problems.

EvoMarket targets scalable, intervention-oriented experiments that require mechanism fidelity for multi-asset, cross-day, and market-specific rules, microstructure fidelity measured against historical LOB summaries, and computational affordability. It also reframes calibration from repeated external black-box loops to a simulator-internal evolution process so that fidelity can be improved within a single run.

\section{System Design}\label{sec:system}

This section describes the design of EvoMarket with an emphasis on high-fidelity, scalable market experimentation. EvoMarket targets two complementary fidelity requirements. First, \emph{mechanism fidelity} requires that institutional trading rules are executed faithfully in settings that extend beyond a single-asset intraday market, including multi-asset interaction, cross-day session structure, and market-specific constraints that shape order submission and price formation. Second, \emph{microstructure fidelity} requires close alignment to historical LOB summaries and order-flow statistics\cite{gould2013lob}. At market scale, these requirements must be met under practical computational budgets and with sufficient observability to support diagnostic analysis.

EvoMarket is also designed for intervention-oriented experiments where a regime $a$ (rules, frictions, or behavior profiles) can be perturbed and outcomes $Y(a)$ can be compared against a control regime under matched seeds and aligned inputs\cite{imbens2024causal,angrist2014mastering}. A central design choice is to move calibration from an external outer loop to an in-run evolution process that exploits observable discrepancy signals, reducing expensive simulator restarts that dominate conventional black-box calibration at scale.

\begin{figure}[pos=t]
  \centering
  \includegraphics[width=1.0\textwidth]{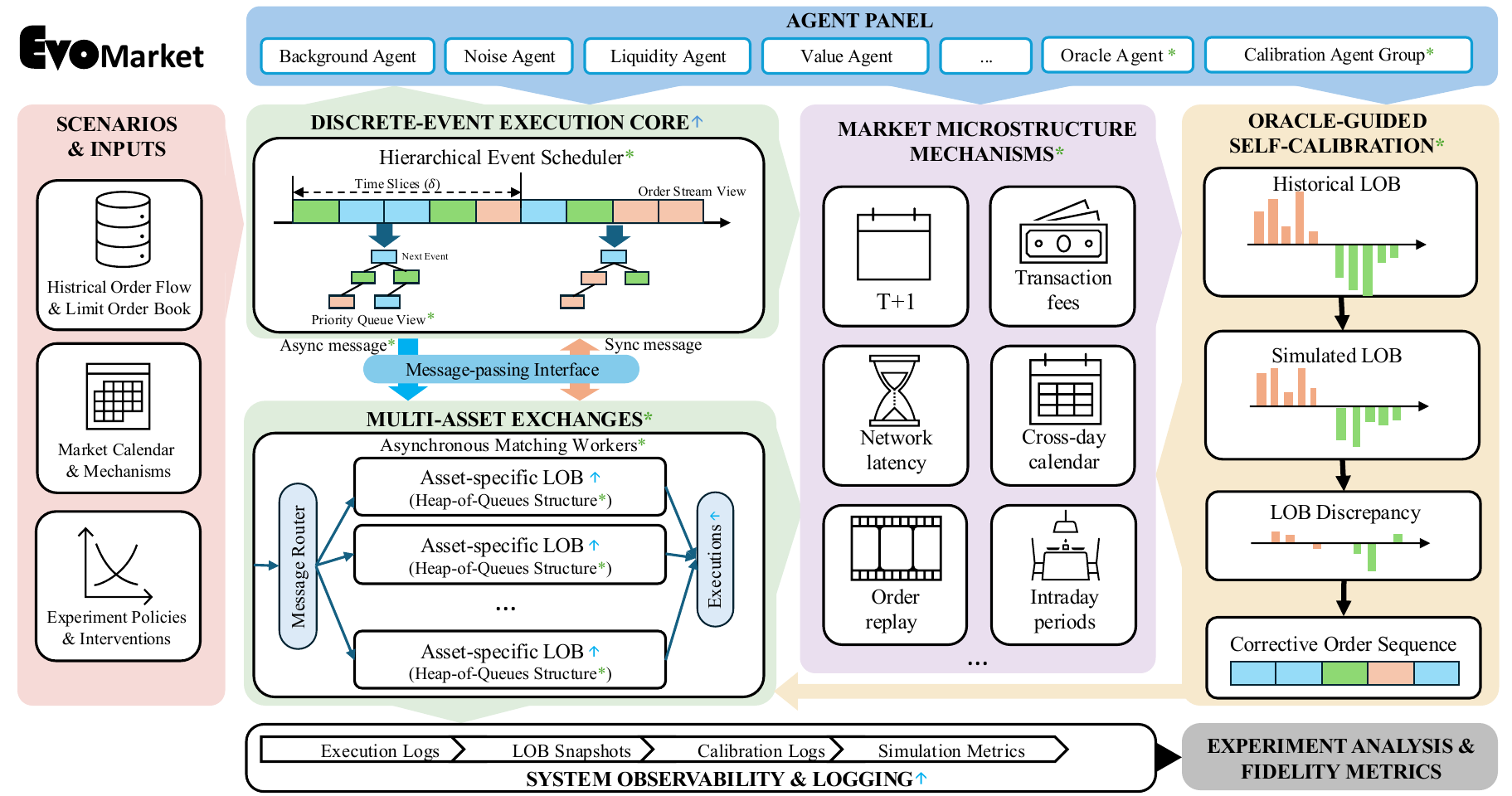}
  \caption{EvoMarket architecture overview. Historical order flow/LOB data, market-calendar mechanisms, and experiment policies drive a discrete-event execution core that routes messages and schedules events, and coordinates multi-asset exchanges and microstructure mechanisms. The Agent Panel hosts heterogeneous financial agent types and provides an interface to extend custom agents. Oracle-guided self-calibration aligns simulated and historical LOBs by converting snapshot discrepancies into corrective order sequences, while the observability layer exports LOB snapshots, execution/calibration logs, and system metrics. Notation: blocks marked with green $^*$ denote EvoMarket-specific designs, blue $^{\uparrow}$ denotes components optimized in EvoMarket, and highlighted background arrows indicate the primary information-flow paths.}\label{fig:sys_arch}
\end{figure}
\FloatBarrier

Fig.~\ref{fig:sys_arch} provides a system-level view of how the discrete-event kernel, multi-asset exchanges, the Oracle, and the instrumentation layer interact to support high-fidelity, repeatable experiments. Formal definitions for markets, orders, and LOBs are given in Section~\ref{sec:prelim}; we build on that notation to describe the design of EvoMarket.

\subsection{Discrete-event Execution Core}
\subsubsection{Event and message model}
EvoMarket follows a discrete-event paradigm\cite{fujimoto1990parallel} in which the simulator state evolves only at event timestamps. We represent an event as $e=(t,\mathrm{type},\mathrm{payload},\pi)$ with execution time $t$, message payload $\mathrm{payload}$, and a deterministic tie-breaker $\pi$ to ensure a total order when timestamps collide, while $\mathrm{type}$ denotes categories such as order arrival/cancellation, exchange response, agent wake-up, session transition, and Oracle triggers. To model propagation delay, a message sent at $t^{\mathrm{send}}$ is received at $t^{\mathrm{recv}}=t^{\mathrm{send}}+\tau^{\mathrm{lat}}$, where $\tau^{\mathrm{lat}}$ is a configurable latency parameter.

\subsubsection{High-throughput order book on the matching critical path}
At market scale, the exchange matching path dominates runtime, so EvoMarket implements $\mathcal{C}$ with a cache-friendly LOB structure. For each asset, the bid and ask sides are maintained as heaps of price levels, and each price level stores a first-in-first-out queue of resting orders. Concretely, let $\mathcal{H}^{\mathrm{ask}}$ be a min-heap keyed by ask price and $\mathcal{H}^{\mathrm{bid}}$ be a max-heap keyed by bid price. Each heap node corresponds to a price level $p$ and points to a linked list (queue) $Q(p)$ that stores orders in arrival order (time priority). We additionally maintain hash maps from price to node so that locating an existing price level is $O(1)$.

This heap-of-queues organization yields $O(1)$ top-of-book access because the best opposite-side price is always at the heap top and the next executable resting order is always at the head of $Q(p)$. For a limit order $o=(\mathrm{id},j,s,\tau,p,v,t)$, the matching loop repeatedly compares the incoming order price against the best opposite-side price and consumes the head order of the best level if executable. Each execution step is $O(1)$ excluding updates, and updates occur only when a price level is created or becomes empty. Inserting a new price level or removing an empty one costs $O(\log P)$ where $P$ is the number of active price levels, while appending to an existing $Q(p)$ is $O(1)$.

\subsubsection{Hierarchical event scheduler under propagation delays}
To schedule heterogeneous events by their receive times under propagation delay, a single global priority queue would require $O(\log N)$ insertion for each event, where $N$ is the total number of pending events, which becomes costly when throughput grows. EvoMarket uses a two-level schedule to reduce this overhead. The timeline is partitioned into coarse time slices of width $\delta$, and each slice maintains a local priority queue ordered by exact timestamps and the tie-breaker $\pi$. An incoming event with time $t$ is first assigned to a slice index $b=\lfloor t/\delta\rfloor$ in $O(1)$ time, then inserted into the local queue of that slice. If the number of events within a slice is $k_b$, the insertion cost becomes $O(\log k_b)$ rather than $O(\log N)$, and in practice $k_b\ll N$ under long-horizon, multi-asset workloads. During execution, the kernel advances slice by slice, draining the local queue and moving to the next non-empty slice, while ordering within each slice is enforced by $(t,\pi)$ so that matched-control experiments remain reproducible.

\subsubsection{Asynchronous per-asset matching for scalable multi-asset execution}
Multi-asset simulation introduces parallelism because CDA matching does not couple assets at the exchange level. EvoMarket exploits this by decoupling order receipt from order processing. When an order-arrival event for asset $j$ becomes due at time $t$, the kernel enqueues it to the matching worker responsible for that asset and continues processing other events, instead of blocking on matching. The worker applies $\mathcal{C}$ to the asset-local LOB state and returns executions, receipts, and LOB updates. To preserve event-time semantics and deterministic replay, EvoMarket commits these results at synchronization points (e.g., at the end of each time slice and immediately before emitting time-aligned snapshots), ensuring that the exchange-side state at time $t$ reflects exactly the set of due events up to $t$.

\subsection{Market Mechanisms for Mechanism Fidelity}
EvoMarket includes a suite of heterogeneous financial agents commonly studied in the agent-based market simulation literature, and it exposes a flexible interface (the \emph{Agent Panel}) to compose and extend custom agent types. This paper does not introduce new trading-agent architectures beyond this interface, and instead defers agent modeling choices to prior work\cite{ehrentreich2008agent,samanidou2007agent}. Instead, the proposed designs focus on  the following market mechanisms.

\subsubsection{Multi-asset exchange and limit order book}
Each asset $j$ is associated with a LOB state $\mathbf{L}^{(j)}_{t}$ within the exchange and a matching operator $\mathcal{C}^{(j)}$ that enforces price-time priority, cancellation, and trade reporting for such asset. The exchange produces structured receipts and trades, and it emits time-aligned LOB snapshots $\{\mathbf{L}^{(j)}_{t}\}_{j=1}^{M}$ as primary observables for microstructure fidelity measurement and self-calibration. A uniform message protocol across assets ensures that a portfolio-level decision $\mathcal{O}_{i,t}$ can be decomposed into asset-specific order-arrival events while preserving causality and deterministic replay.

\subsubsection{Cross-day calendar and session transitions}
Cross-day simulation requires explicit calendars and session-specific mechanisms. For each trading day $d$, EvoMarket defines a calendar $\mathcal{K}_d$ as an ordered list of sessions $\{(s_k,[t_{d,k}^{\mathrm{start}},t_{d,k}^{\mathrm{end}}))\}_{k=1}^{K_d}$, including a preopen call auction, continuous trading, an intraday break, and an end-of-day clearing stage. These transitions are represented as explicit kernel events so that they can be replayed and perturbed under different regimes $a$.
For the A-share calendar used in this paper, we set $k=1,\ldots,K_d$ with $K_d=4$ stages corresponding to (i) preopen call auction, (ii) continuous trading, (iii) intraday break, and (iv) end-of-day clearing.

This design is necessary for mechanism fidelity because a substantial fraction of real-world trading concentrates around session boundaries, especially the opening and closing periods where liquidity aggregation and information arrival are strongest. If a simulator omits preopen auctions or end-of-day clearing, it cannot faithfully support strategy evaluation and policy experiments whose outcomes depend on these phases.

In EvoMarket, the preopen call auction is modeled as an operator $\mathcal{A}^{(j)}$ that aggregates preopen orders for asset $j$ and computes a clearing price $p^{(j)}_{d,\mathrm{open}}$ and executions at the session transition as
\begin{equation}
(p^{(j)}_{d,\mathrm{open}}, \mathcal{E}^{(j)}_{d,\mathrm{open}})=\mathcal{A}^{(j)}(\mathcal{O}^{(j)}_{d,\mathrm{preopen}}),
\end{equation}
where $\mathcal{O}^{(j)}_{d,\mathrm{preopen}}$ denotes the set of preopen orders received in the auction window and $\mathcal{E}^{(j)}_{d,\mathrm{open}}$ denotes the resulting executions. During the intraday break, matching is suspended while the kernel continues to advance time and to schedule the next session boundary. At end-of-day, EvoMarket performs a clearing event that finalizes the day, cancels any unfilled day orders as configured, and triggers settlement-state transitions used by market-specific constraints.

\subsubsection{Market-specific constraints}
EvoMarket implements market-specific institutional constraints of the China A-share market that materially affect feasible order flow and price formation\cite{madhavan2000market}. First, daily price limits restrict executable prices to a band around a reference price. Let $p^{(j)}_{d,\mathrm{ref}}$ be the reference price of asset $j$ on day $d$ (typically the previous close) and let $\eta$ be the limit ratio. EvoMarket enforces
\begin{equation}
p^{(j)}_{d,\min}=(1-\eta)p^{(j)}_{d,\mathrm{ref}},\quad p^{(j)}_{d,\max}=(1+\eta)p^{(j)}_{d,\mathrm{ref}},
\end{equation}
and rejects or truncates limit orders with prices outside $[p^{(j)}_{d,\min},p^{(j)}_{d,\max}]$ according to exchange rules. This constraint shapes liquidity provision and contributes to discontinuities around the limit boundaries.

Second, the T+1 settlement rule constrains same-day inventory usage. We decompose holdings into available and pending components, $\mathbf{h}_{i,t}=\mathbf{h}^{\mathrm{avail}}_{i,t}+\mathbf{h}^{\mathrm{pend}}_{i,t}$, where $\mathbf{h}^{\mathrm{avail}}_{i,t}$ denotes currently sellable inventory and $\mathbf{h}^{\mathrm{pend}}_{i,t}$ denotes shares purchased on day $d$ that are not sellable until the next settlement. For a sell order $o=(\mathrm{id},j,\mathrm{sell},\tau,p,v,t)$ submitted on day $d$, EvoMarket enforces the feasibility constraint
\(v \le h^{\mathrm{avail}}_{i,t}(j),\) and at the end-of-day clearing event it updates $h^{\mathrm{avail}}_{i,t}(j)\leftarrow h^{\mathrm{avail}}_{i,t}(j)+h^{\mathrm{pend}}_{i,t}(j)$ and resets $h^{\mathrm{pend}}_{i,t}(j)\leftarrow 0$ for all $j$. These constraints are critical for reproducing realistic short-horizon inventory dynamics and for evaluating strategies and interventions in A-share settings.

\subsubsection{Cross-asset linkage interface}
Cross-asset experiments require coupling mechanisms beyond independent per-asset LOBs. EvoMarket therefore exposes portfolio-level primitives and cross-asset information queries. At the agent layer, a decision policy may depend on a multi-asset state summary $x_t=\psi(\{\mathbf{L}^{(j)}_{t}\}_{j=1}^{M})$ and produce a batch order set $\mathcal{O}_{i,t}$ that places orders on multiple assets simultaneously, reflecting realistic portfolio trading and hedging behavior. At the system layer, EvoMarket provides interfaces for shared risk factors, cross-asset constraints, and intervention-driven shocks, enabling controlled studies of correlation structure and shock propagation. The goal is to make linkage an explicit, configurable mechanism so that observed cross-asset effects can be attributed to specific portfolio constraints, information channels, or institutional frictions.

\subsection{Self-calibration as Simulator-internal Evolution}
\subsubsection{Oracle-guided corrective order synthesis}
EvoMarket performs self-calibration by synthesizing \emph{corrective orders} that reduce microstructure discrepancy while minimally perturbing the ongoing simulation. The key observation is that the discrepancy between the simulated LOB and the historical reference can be interpreted as a missing order-flow component. In particular, given a simulated snapshot $\mathbf{L}^{(j)}_{t}$ and an Oracle-provided target snapshot $\tilde{\mathbf{L}}^{(j)}_{t'}$ for asset $j$ at a future time $t'>t$, the calibration objective is to construct a short corrective order sequence $\mathcal{O}^{(j)}_{\mathrm{cal}}(t\!\rightarrow\!t')$ such that applying CDA yields a state close to the target:
\begin{equation}
\tilde{\mathbf{L}}^{(j)}_{t'} \approx \mathcal{C}^{(j)}\!\left(\mathbf{L}^{(j)}_{t}, \mathcal{O}^{(j)}_{\mathrm{cal}}(t\!\rightarrow\!t')\right),
\end{equation}
which aligns with a minimal-order-difference principle. Among all order sequences that can transform one snapshot to another under CDA, the shortest sequence is preferred, thereby achieving a targeted correction with minimal perturbation.

To formalize this notion, we overload $\mathcal{C}(\mathbf{L},\mathcal{O})$ to denote the resulting snapshot after applying an ordered sequence of orders $\mathcal{O}=(o^{1},\ldots,o^{m})$ under CDA. Given two snapshots $\mathbf{L}^{a}$ and $\mathbf{L}^{b}$, the \emph{minimal order difference} is defined as the shortest order sequence that transforms $\mathbf{L}^{a}$ into $\mathbf{L}^{b}$:
\begin{equation}
\mathcal{O}^{*}(\mathbf{L}^{a}\rightarrow \mathbf{L}^{b})
=\arg\min_{\mathcal{O}}|\mathcal{O}|
\quad\mathrm{s.t.}\quad
\mathbf{L}^{b}= \mathcal{C}(\mathbf{L}^{a},\mathcal{O}),
\label{eq:min_order_diff}
\end{equation}
where $|\mathcal{O}|$ is the sequence length. When the constraint is infeasible due to aggregation (snapshots discard queue-level details), Eq.~\eqref{eq:min_order_diff} can be relaxed to match $\mathbf{L}^{b}$ up to a small discrepancy under a chosen distance over snapshot features. In this paper, we use level-$1$ to level-$L$ best bid/ask prices and aggregated depths as the snapshot features targeted by calibration. EvoMarket uses this principle to construct corrective orders that reduce microstructure discrepancies with minimal perturbation.

To make this construction tractable in-run, EvoMarket introduces an \emph{Oracle} that provides calibration agents with a reference snapshot at specified future timestamps. Let $\mathcal{T}_{\mathrm{rec}}=\{t_k\}$ be the set of LOB recording timesteps implied by the snapshot frequency. For each asset $j$ and each checkpoint $t_k$, a calibration agent queries the Oracle for the reference LOB at the next recording time $t_{k+1}$:
\begin{equation}
\tilde{\mathbf{L}}^{(j)}_{t_{k+1}} = \mathsf{Oracle}(j, t_{k+1}; a),
\end{equation}
where $a$ denotes the current experimental regime including user interventions. The calibration agent then computes a level-wise gap signal $\mathbf{G}^{(j)}_{k}$ between the simulated and reference LOBs (differences of prices and depths across levels $1\ldots L$ on both sides), and passes it to a calibration agent group that generates corrective orders using a greedy heuristic.

This checkpoint-level control flow is triggered only at recording times, and the corrective orders are matched immediately before logging, enabling in-run fidelity improvement under a fixed compute budget.

\subsubsection{Greedy calibration algorithm}
At each checkpoint $t_k$, calibration agents operate right before the LOB is recorded at $t_{k+1}$. For each asset $j$, they compare the simulated LOB $\mathbf{L}^{(j)}_{t_k}$ with the Oracle snapshot $\tilde{\mathbf{L}}^{(j)}_{t_{k+1}}$ and compute a gap tensor $\mathbf{G}^{(j)}_k$ (e.g., per-level depth differences on both sides). The calibration agent group then synthesizes a corrective order list by greedily addressing the largest gaps first, producing a sequence $\mathcal{O}^{(j)}_{\mathrm{cal}}$ that is submitted to the exchange and matched by $\mathcal{C}^{(j)}$. This approach approximates the shortest corrective sequence in Eq.~\eqref{eq:min_order_diff} while remaining fast enough for market-scale runs.

\begin{algorithm}[!t]
\caption{Oracle-guided online self-calibration at checkpoint $t_k$}
\label{alg:self_calib}
\begin{algorithmic}[1]
\REQUIRE Recording times $\mathcal{T}_{\mathrm{rec}}=\{t_k\}$, assets $j=1,\ldots,M$, exchanges $\{\mathcal{C}^{(j)}\}$, Oracle $\mathsf{Oracle}(\cdot)$
\FOR{each checkpoint $t_k\in\mathcal{T}_{\mathrm{rec}}$}
\FOR{each asset $j=1,\ldots,M$ in parallel}
\STATE Query Oracle: $\tilde{\mathbf{L}}^{(j)}_{t_{k+1}} \leftarrow \mathsf{Oracle}(j,t_{k+1};a)$
\STATE Compute gap: $\mathbf{G}^{(j)}_{k} \leftarrow \mathsf{Gap}(\mathbf{L}^{(j)}_{t_k}, \tilde{\mathbf{L}}^{(j)}_{t_{k+1}})$
\STATE Prioritize gaps: $\mathcal{G}^{(j)} \leftarrow \mathsf{SortByMagnitude}(\mathbf{G}^{(j)}_{k})$
\STATE Synthesize corrective orders (greedy): $\mathcal{O}^{(j)}_{\mathrm{cal}} \leftarrow \mathsf{GreedySynthesize}(\mathcal{G}^{(j)})$
\STATE Submit $\mathcal{O}^{(j)}_{\mathrm{cal}}$ to exchange $j$; match via $\mathcal{C}^{(j)}$
\ENDFOR
\STATE Advance kernel to $t_{k+1}$; record $\{\mathbf{L}^{(j)}_{t_{k+1}}\}_{j=1}^{M}$
\ENDFOR
\end{algorithmic}
\end{algorithm}

To make the comparison between in-run self-calibration and outer-loop calibration more explicit, we analyze both methods under a common baseline. Consider a base calibration problem with one asset, a fixed one-hour window, and a fixed agent population, and treat the wall-clock cost of one full simulator run as a constant baseline. In-run self-calibration operates on bounded-depth snapshot summaries (at most 10 LOB levels in this paper) and performs one discrepancy-to-order construction at each checkpoint without searching a parameter space. Therefore, for the base problem, the calibration overhead is a constant-factor augmentation of one run and can be treated as $\mathcal{O}(1)$ in terms of simulation runs.

Outer-loop calibration instead treats the simulator as a black-box objective $J(\theta)$ and searches over a $d$-dimensional parameter vector $\theta\in\mathbb{R}^{d}$, where $d$ typically scales with the number of calibrated agents. Let $N$ denote the number of full-run objective evaluations. For PSO with swarm size $S$ and $I$ iterations, $N=SI$. Since each evaluation requires a full run over the same one-hour window, the base-problem wall-clock scales as $\mathcal{O}(N)$ in terms of simulation runs.

The critical difference is how $N$ scales with the effective dimension $d$ when comparable accuracy is required. Suppose the acceptable parameter set corresponds to an $\varepsilon$-neighborhood in a bounded $d$-dimensional search region. Black-box methods that rely on sampling this region face a covering-number barrier, since resolving the space at granularity $\varepsilon$ requires on the order of $(1/\varepsilon)^{d}$ regions. This implies that the evaluation count needed to reliably reach an acceptable region can grow at least exponentially in $d$, yielding a curse-of-dimensionality effect as the calibrated agent population and parameterization expand.

When extending calibration to $M$ assets, in-run self-calibration continues to operate through per-asset discrepancy signals and per-asset corrective-order synthesis, so wall-clock grows approximately linearly in $M$ and can parallelize across assets, i.e., $\mathcal{O}(M)$ under the same normalization. For outer-loop calibration, if assets are calibrated independently, the cost becomes $\mathcal{O}(MN(d))$ because each asset requires its own sequence of full-run evaluations. If calibration is performed jointly to account for cross-asset coupling, the effective dimension grows with $M$, and the evaluation count can inherit an exponential dependence on dimension, leading to prohibitive overall wall-clock.

\subsubsection{Intervention-induced ambiguity in Oracle information}
When an intervention injects additional orders or rule changes, the counterfactual future LOB is no longer uniquely determined by the historical reference. EvoMarket models this ambiguity by perturbing the Oracle-provided snapshot with a random error whose variance increases with the intervention magnitude. Let $\mathcal{I}_{k}$ denote the set of intervention orders injected during the interval $(t_k, t_{k+1}]$ and define its size as the total intervened volume
\begin{equation}
V(\mathcal{I}_{k}) = \sum_{o\in \mathcal{I}_{k}} v(o).
\end{equation}
The Oracle returns a noisy reference snapshot
\begin{equation}
\tilde{\mathbf{L}}^{(j)}_{t_{k+1}} = \mathbf{L}^{(j),\star}_{t_{k+1}} + \boldsymbol{\epsilon}^{(j)}_{k},
\quad
\boldsymbol{\epsilon}^{(j)}_{k} \sim \mathcal{N}\!\left(\mathbf{0}, \sigma^2(V(\mathcal{I}_{k}))\mathbf{I}\right),
\label{eq:oracle_noise}
\end{equation}
where $\sigma^2(\cdot)$ is a monotone function, for example $\sigma^2(V)=\sigma_0^2+\alpha V$, and the noise is applied to the level-$1$ to level-$L$ bid/ask prices and depths used by calibration. Intuitively, stronger interventions induce larger deviations from the historical trajectory, and the \emph{Oracle} provides only an uncertain reference consistent with this fact.

Because calibration operates at each recording checkpoint and explicitly targets the LOB gap, it keeps the snapshot-level error between simulated and historical trajectories small by construction, subject to the intervention ambiguity in Eq.~\eqref{eq:oracle_noise}. Section~\ref{sec:experiments} evaluates the resulting fidelity improvement and the computational cost of this in-run evolution relative to external iterative calibration procedures.

\section{Experiments}\label{sec:experiments}

This section evaluates EvoMarket along the core claims stated in Section~\ref{sec:intro} and Section~\ref{sec:system}. We organize the results into three groups, namely (i) \emph{mechanism fidelity} (correct execution under replay and multi-asset dependence), (ii) \emph{microstructure fidelity} improved via in-run self-calibration, and (iii) \emph{scalable efficiency} (throughput, market-breadth scaling, and ablations). Additional diagnostics are available in the supplementary material.

\subsection{Experimental setup}
\textbf{Data.} We use China A-share historical order-flow records and aligned LOB snapshots. Unless otherwise noted, time-series are aligned at the sampling frequency of $3$ seconds within the same market window.

\textbf{Baselines.} When applicable, we compare against representative simulators ABIDES\cite{byrd2019abides} and MAXE\cite{belcak2022maxe}.

\textbf{Metrics.} We report (i) snapshot-level microstructure errors (LOB-MSE, mid-price path alignment, and distributional diagnostics), (ii) system-level performance (throughput, wall-clock, and memory scaling), and (iii) linkage-level observables for multi-asset settings (cross-asset correlation structure). For external calibration baselines, we report best-so-far price MSE versus calibration wall-clock, where PSO runs up to one hour to match the one-hour target window (3-second snapshots).

\subsection{Simulation Studies on Mechanism fidelity}
\subsubsection{Historical replay of the matching mechanism}
We first validate the correctness foundation of matching, cancellation handling, and event-time alignment by replaying historical order flow, recording simulated snapshots, and comparing them with historical references. Note that the replay data used here starts from continuous trading (09:30), so pre-open call-auction dynamics are intentionally omitted in this replay experiment.

\begin{figure}[pos=htbp]
  \centering
  \includegraphics[width=0.60\linewidth]{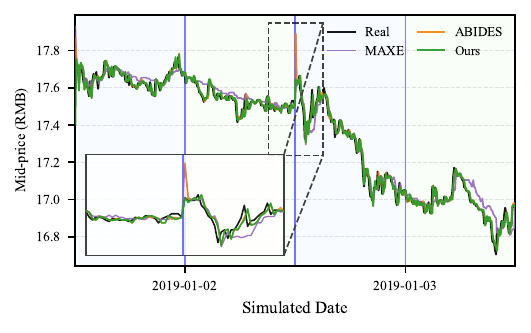}
  \caption{Mid-price alignment under historical order replay for Jan 2--3 (1-minute sampling; pre-open, lunch-break, and overnight intervals omitted). Light blue and light green background bands indicate morning and afternoon sessions; blue vertical lines indicate session boundaries (lunch and day transitions). The inset zooms into the close-to-open transition.}\label{fig:replay_fidelity}
\end{figure}
\FloatBarrier

Across these two trading days, EvoMarket closely tracks the historical mid-price path and remains competitive with representative baselines. The inset highlights the close-to-open transition, where day-level state handling is practically important. Although the replay begins at continuous trading (09:30) and does not replay pre-open auctions, the initial LOB at 09:30 is the output of the previous close and pre-open order aggregation in the real market, so a simulator that resets the LOB can deviate immediately at market open and propagate errors into the rest of the day. EvoMarket treats day transitions as explicit kernel events and supports consistent day-start initialization, which yields a stable continuation across days, whereas ABIDES and MAXE exhibit larger transient mismatch around the boundary. This difference is consequential rather than cosmetic, since boundary-induced drift can bias both strategy evaluation and rule-intervention analysis.

\subsubsection{Agent order-space coverage}
We analyze an order trace from a single simulation run with $1.27$M recorded orders to assess behavioral diversity and to verify that the mechanism layer supports heterogeneous financial agents. EvoMarket does not introduce new agent models beyond the Oracle and the calibration agent group; instead, we reuse representative agent implementations from prior work and expose a unified interface for extending custom agent types (the \emph{Agent Panel}). We focus on \emph{limit orders} and visualize each agent type's order distribution in the $(\Delta_{\mathrm{tick}},q)$ plane, where $\Delta_{\mathrm{tick}}=(p-m)/\Delta p$ is the tick offset from the contemporaneous mid-price $m=(p^{\mathrm{bid}}+p^{\mathrm{ask}})/2$, and $\Delta p$ denotes the tick size (RMB~0.01 for China A-shares). The vertical axis reports order size in lots (the minimum trading unit in A-shares). We plot Gaussian-smoothed density overlays (darker indicates higher density) under log-scaled axes (symlog for $\Delta_{\mathrm{tick}}$ and log for $q$).

\begin{figure}[pos=htbp]
  \centering
  \includegraphics[width=0.60\linewidth]{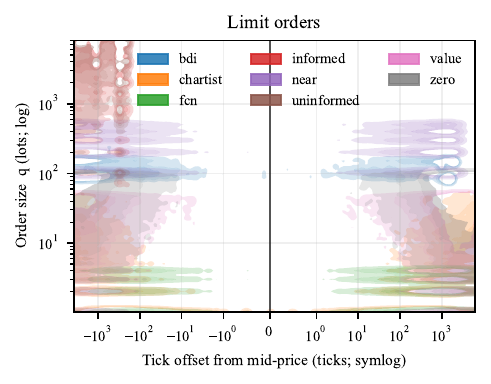}
  \caption{Agent order-space coverage for limit orders (density overlays by agent type). The horizontal axis is the tick offset from mid-price $\Delta_{\mathrm{tick}}=(p-m)/\Delta p$ (symlog); the vertical axis is order size $q$ in lots (log).}\label{fig:g1_space}
\end{figure}
\FloatBarrier

Fig.~\ref{fig:g1_space} provides a compact diagnostic of whether the simulator can express realistic order-flow heterogeneity. Orders spread across both sides of the mid-price and span several orders of magnitude in size, indicating broad coverage of the limit-order space rather than a narrow, hand-tuned operating regime. This matters because calibration and intervention studies are only meaningful if the simulator can represent the regions of the order space that dominate real trading. When the agent population is overly concentrated near a single price offset or size scale, a good fit can be an artifact of model rigidity rather than genuine mechanism fidelity. From an experimental perspective, the resulting coverage also provides a concrete basis for assigning corrective orders to heterogeneous agent groups during self-calibration, since different agent types naturally occupy different parts of the order space.

\subsubsection{Cross-asset linkages}
To demonstrate market-wide experimentation beyond independent per-asset replay, we evaluate whether EvoMarket can produce non-trivial cross-asset dependence structure within a single multi-asset simulation. In real markets, assets are coupled through common risk factors, sector and industry linkages, and correlated order flow, so a simulator that treats assets as independent processes cannot support portfolio-level studies that require cross-sectional patterns and joint dynamics.

\textbf{Cross-asset dependence.} We run ABIDES 20 times with different random seeds and treat the 20 resulting mid-price series as 20 single assets. We compare this against one EvoMarket run with 20 assets. In both cases, we compute cross-asset correlations from 1-min mid-price log returns. Specifically, let $m_i(t)$ denote the mid-price of asset $i$ at minute $t$, where $m_i(t)=\big(p_i^{\mathrm{bid}}(t)+p_i^{\mathrm{ask}}(t)\big)/2$. The 1-min log return is
\(r_i(t) \;=\; \log m_i(t) - \log m_i(t-1).\)
We then estimate the correlation matrix $\mathbf{C}\in\mathbb{R}^{N\times N}$ with entries
\(C_{ij} \;=\; \mathrm{corr}\!\left(r_i, r_j\right), \qquad i,j\in\{1,\ldots,N\},\)
where $N=20$ and $\mathrm{corr}(\cdot,\cdot)$ denotes the Pearson correlation computed over the aligned return samples.

\begin{figure}[pos=htbp]
  \centering
  \includegraphics[width=0.60\linewidth]{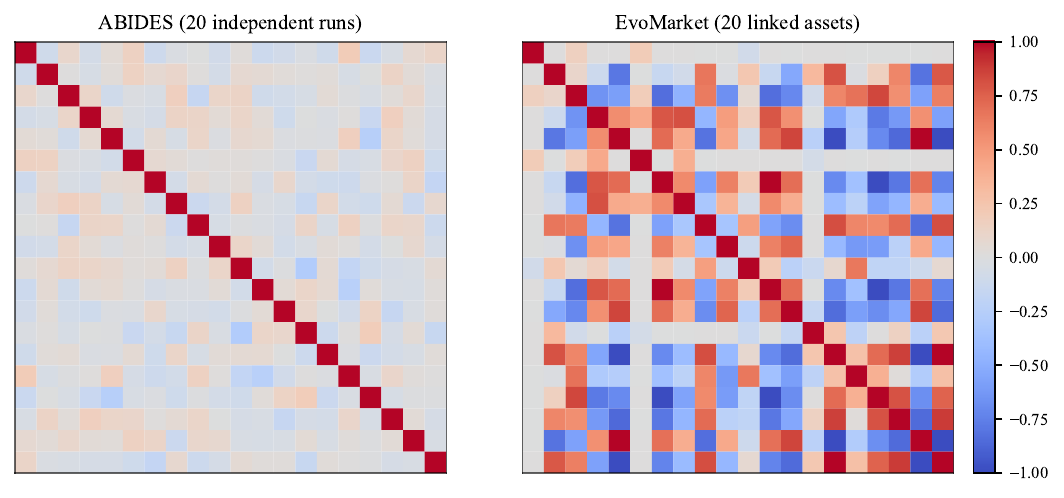}
  \caption{Cross-asset correlation heatmaps computed from 1-min mid-price log returns (ABIDES, 20 independent runs treated as 20 assets; EvoMarket, single linked 20-asset run).}\label{fig:crossasset_corr}
\end{figure}
\FloatBarrier

Fig.~\ref{fig:crossasset_corr} shows that the baseline composed of independent ABIDES runs is close to uncorrelated, with off-diagonal correlations concentrated near zero. By contrast, EvoMarket produces a structured dependence pattern with both positive and negative co-movement, including pronounced correlated and anti-correlated groups. This result is not intended to match any specific empirical correlation matrix; rather, it demonstrates that cross-asset coupling can be expressed as a first-class mechanism within a single simulation run instead of being approximated by multiple independent single-asset runs.

This capability has direct practical implications. Cross-sectional research and alpha factor mining require realistic joint dynamics as a substrate for hypothesis generation and stress testing, while portfolio construction and execution depend on how shocks and trading pressure propagate across assets. A simulator that can generate structured dependence within one run can therefore support end-to-end evaluation of multi-asset strategies under controlled interventions. Moreover, when combined with high-throughput execution and replay-level observability, such a simulator can reduce reliance on lengthy forward paper-trading cycles by enabling reproducible, multi-month historical replay studies with portfolio-level agents under consistent mechanism and instrumentation.

\subsubsection{Case study on intervention evaluation}
We next demonstrate intervention-ready evaluation with an event-study style plot using mid-price only. We first generate a deterministic baseline by fixing the simulator seed and running a simple quote-updater process that periodically posts and cancels small limit orders around the prevailing mid-price, yielding a non-trivial baseline trajectory. We then inject an intervention at a fixed time and re-run calibration using an oracle that provides \emph{noisy} post-event LOB targets (Gaussian perturbations around the shocked LOB), inducing variation across calibrated post-event trajectories.

\begin{figure}[pos=htbp]
  \centering
  \includegraphics[width=0.98\linewidth]{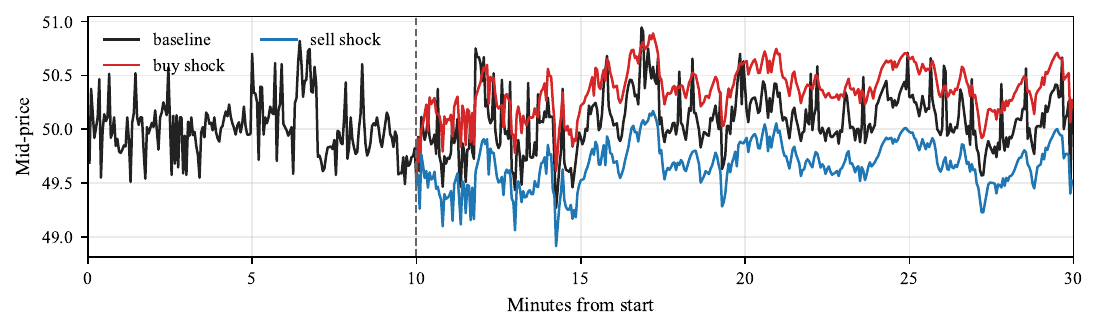}
  \caption{Event-study style intervention evaluation (mid-price only). A step-jump intervention is injected at a fixed time to mimic an aggressive buy/sell sweep. The baseline curve is deterministic under a fixed seed; each intervention direction is repeated 10 times and shown as a mean trajectory after the event.}\label{fig:event_study}
\end{figure}

We illustrate one intervention primitive with a positive and negative direction, namely a \emph{step jump} that mimics an aggressive buy or sell sweep that instantaneously shifts the clearing region. For each direction, we run 10 calibrations and plot the mean trajectory after the event time. Fig.~\ref{fig:event_study} aligns outcomes in event time and visualizes counterfactual trajectories $Y(a)$ under perturbed regimes $a$ against the matched-seed control $Y(a_0)$. This matters for intervention-oriented studies because it isolates the effect of a specified shock and enables sensitivity analysis under oracle ambiguity through the dispersion across repeated calibrations.

\FloatBarrier

\subsection{Simulation Studies on Microstructure fidelity}
\subsubsection{In-run self-calibration under fixed compute budgets}
We evaluate calibration on one hour of market data at 3-seconds sampling frequency. EvoMarket performs self-calibration in-run and therefore produces a calibrated trajectory in a single simulator execution. By contrast, ABIDES and MAXE treat calibration as an external optimization loop that repeatedly re-runs the simulator over the full one-hour window. To reflect a practical real-time constraint, we calibrate ABIDES and MAXE with PSO starting from their default parameterizations and limit the total calibration wall-clock to one hour, matching the length of the calibration window.

\begin{figure}[pos=htbp]
  \centering
  \includegraphics[width=0.50\linewidth]{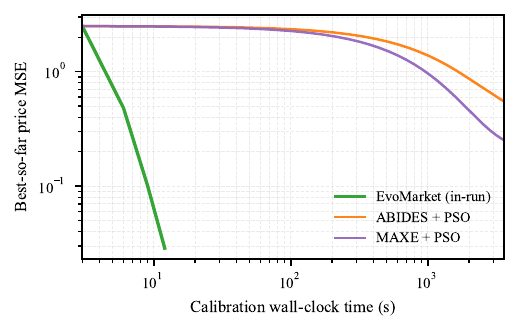}
  \caption{Calibration efficiency on one hour of 3-second snapshots. Best-so-far price MSE versus wall-clock time is shown in log-log scale. EvoMarket performs in-run self-calibration and ends after one run, while ABIDES+PSO and MAXE+PSO run an external optimizer for one hour (3{,}600 seconds), matching the calibration window length.}\label{fig:calib_cost}
\end{figure}

To contextualize convergence speed, Fig.~\ref{fig:calib_cost} reports best-so-far price MSE versus wall-clock time with both axes in log scale. EvoMarket completes one calibrated run in about 12 seconds and achieves the lowest MSE within this single execution, while ABIDES+PSO and MAXE+PSO consume the full one-hour wall-clock yet remain substantially higher. In this benchmark, in-run self-calibration yields about $300\times$ less wall-clock time and about $10 \times$ lower MSE. This separation matters because calibration becomes a minor in-run overhead rather than the dominant cost driver, making microstructure-level alignment feasible in repeated experiments and in market-scale settings where external restarts would be prohibitive.

\FloatBarrier

\begin{figure}[pos=htbp]
  \centering
  \includegraphics[width=0.6\linewidth]{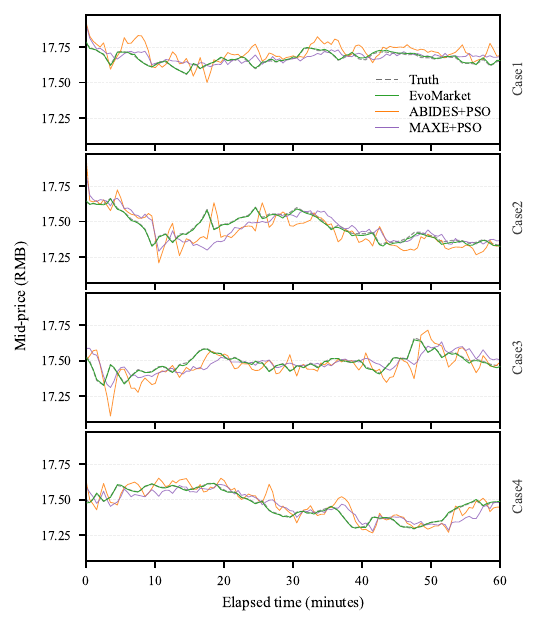}
  \caption{One-hour calibration result on 3-second snapshots. Four independent 60-minute cases are visualized at 30-second sampling for readability. Mid-price is measured in RMB. EvoMarket remains close to truth, while ABIDES+PSO and MAXE+PSO show larger residual deviations under the same one-hour wall-clock limit.}\label{fig:calib_windows}
\end{figure}

Fig.~\ref{fig:calib_windows} visualizes four independent 60-minute cases over the same target interval. EvoMarket closely tracks the historical mid-price path, while ABIDES+PSO and MAXE+PSO still show visible residual mismatch under the same one-hour wall-clock limit. This result indicates that external black-box calibration does not reach comparable microstructure accuracy within the time scale required for practical use.

\FloatBarrier

\begin{figure}[pos=htbp]
  \centering
  \includegraphics[width=0.7\linewidth]{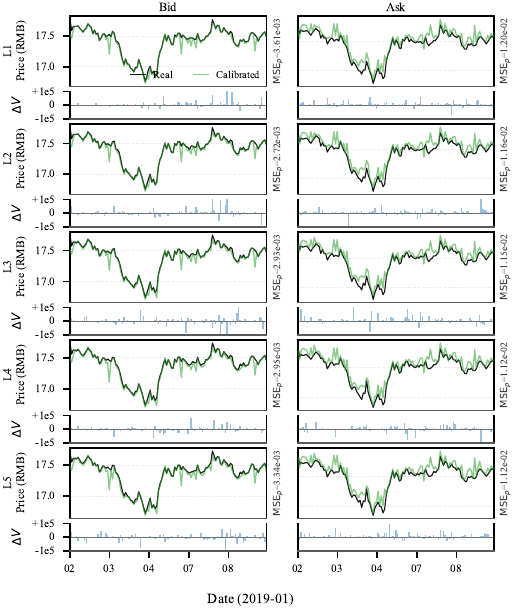}
  \caption{Calibrated vs.\ real LOB at levels 1--5 (bid/ask) over five days (10-minute sampling; pre-open, lunch break, and overnight intervals omitted). Prices (RMB) are shown as lines and $\Delta V$ (shares; calibrated $-$ real) is shown as bars (clipped to $\pm 1e5$). Each panel annotates price MSE for the corresponding level. Date ticks indicate day-of-month in January 2019.}\label{fig:calibration}
\end{figure}

Fig.~\ref{fig:calibration} further shows that self-calibration reduces snapshot-level discrepancies across multiple depth levels within a single run. Across levels 1--5, the calibrated bid and ask prices closely overlap with the historical trajectory, and the price MSE in each panel is typically at the $10^{-3}$ scale, corresponding to root-mean-squared errors on the order of a few $\times 10^{-2}$ RMB (tick size RMB~0.01). Therefore, self-calibration yields tick-level LOB alignment beyond mid-price tracking, which is necessary for microstructure diagnostics and for intervention analyses that depend on depth dynamics.

\FloatBarrier

\subsection{Scalable efficiency and ablations}
We next validate computational tractability at market scale by stressing the event-driven engine under increasing input order rates and evaluating market-breadth scaling by increasing the number of stocks while controlling per-stock workload.

\begin{figure}[pos=htbp]
  \centering
  \includegraphics[width=0.5\linewidth]{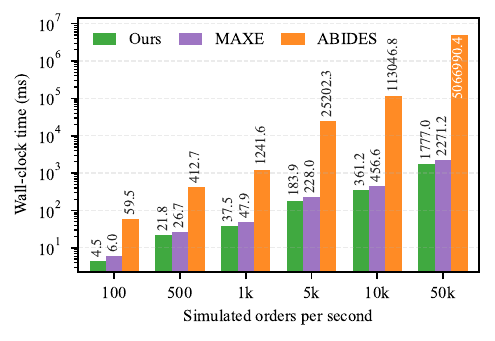}
  \caption{Single-core throughput stress test. Wall-clock time to process a given simulated order rate (orders per second) is reported under increasing load (log-scale x-axis). Lower is better. Values are mean execution times (ms).}\label{fig:b0_single_core}
\end{figure}
\FloatBarrier

\begin{figure}[pos=htbp]
  \centering
  \includegraphics[width=0.6\linewidth]{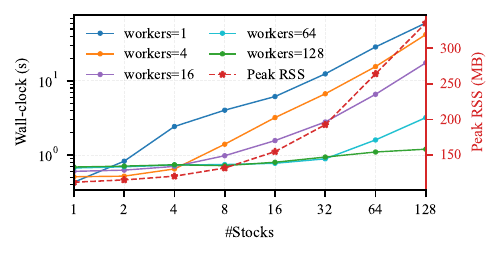}
  \caption{Market-breadth scaling under fixed per-stock order density. We increase the number of stocks while keeping the per-stock order injection rate fixed at 200 orders per second for a 10-second simulated window, and report wall-clock time (left axis, log scale) and peak RSS (right axis). Curves show worker settings 1/4/16/64/128.}\label{fig:b2_scaling}
\end{figure}

\begin{table}[pos=htbp]
  \centering
  \caption{Ablation on engine efficiency. We vary exchange-side parallelism (Workers), asynchronous mechanism (Async), and logging overhead.}
  \label{tab:b3_ablation}
  \scriptsize
  \setlength{\tabcolsep}{3pt}
  \resizebox{0.6\textwidth}{!}{%
    \begin{tabular}{lrrrrrrr}
      \toprule
      Config & Workers & Async & LOB\_freq & Main\_log & Throughput & Time(s) & Log\_size(MB) \\
      \midrule
      C0 & 0 & F & 3s & T & 79{,}252 & 113.56 & 129.4 \\
      C1 & 2 & F & 3s & T & 127{,}779 & 70.43 & 129.4 \\
      C2 & 2 & T & 3s & T & 135{,}864 & 66.24 & 129.4 \\
      C3 & 2 & T & 30s & T & 223{,}875 & 40.20 & 13.1 \\
      C4 & 2 & T & 30s & F & \textbf{238{,}224} & \textbf{37.78} & \textbf{2.4} \\
      \bottomrule
    \end{tabular}%
  }
\end{table}
\FloatBarrier

Fig.~\ref{fig:b0_single_core} shows that EvoMarket sustains high input order rates with substantially lower wall-clock cost than representative baselines in a single-core setting, especially at high order rates. Fig.~\ref{fig:b2_scaling} shows a clear scalability pattern under fixed per-stock order density. Wall-clock remains relatively flat when the number of stocks is within the available worker budget and increases approximately linearly once the market breadth exceeds this budget, while peak memory grows with the number of LOBs. Here, peak RSS refers to the peak resident set size of the simulator process (physical memory footprint) measured during the run.

\textbf{Ablation.} Table~\ref{tab:b3_ablation} quantifies the contribution of each engineering choice. Turning on exchange workers (C0$\rightarrow$C1) raises throughput from 79,252 to 127,779 orders per second (+61.2\%) and reduces wall-clock from 113.56 seconds to 70.43 seconds (-38.0\%). Adding asynchronous queries (C1$\rightarrow$C2) brings a further +6.3\% throughput gain and -5.9\% time reduction. Lowering LOB logging frequency from 3-seconds to 30-seconds (C2$\rightarrow$C3) yields the largest I/O win: throughput rises to 223,875 (+64.8\%) while log size drops from 129.4MB to 13.1MB (about 9.9$\times$ smaller). Disabling the main log (C3$\rightarrow$C4) gives an additional +6.4\% throughput gain and reduces log size to 2.4MB (about 5.5$\times$ smaller than C3). Overall, exchange-side parallelism and reduced logging frequency dominate the throughput gains, while asynchronous queries provide a smaller incremental improvement.

\textbf{Experimental takeaway.} This work responds to the two questions raised in Section I as follows. For Q1 (\emph{for fidelity and scalability}), EvoMarket simultaneously preserves mechanism realism (replay consistency, cross-asset linkage, and intervention responses) and achieves tight microstructure alignment after calibration. For Q2 (\emph{for computational efficiency}), EvoMarket reaches the target calibration accuracy within a practical real-time budget and retains scalable execution, with nearly linear wall-clock growth as market breadth increases.

\section{Conclusion}\label{sec:conclusion}

This paper presented \textit{EvoMarket}, a high-fidelity and scalable financial market simulator designed for intervention-oriented computational experiments. EvoMarket unifies (i) mechanism fidelity for multi-asset and cross-day execution with market-specific institutional rules, (ii) microstructure fidelity aligned to historical LOB observables via an Oracle-guided in-run self-calibration loop, and (iii) scalable discrete-event engineering that keeps market-wide studies computationally tractable.

Our experiments validate the core claims of the paper: EvoMarket closely tracks historical LOB dynamics under replay and improves microstructure fidelity within a single run via Oracle-guided self-calibration; it sustains high throughput and exhibits predictable wall-clock and memory scaling with market breadth; and it supports market-wide evaluation including cross-asset linkages and event-study style intervention analysis. Order-trace analysis additionally indicates broad agent order-space coverage.

Future work will focus on stronger self-calibration algorithms beyond the current heuristic, broader evaluation across assets and time spans with standardized benchmarks, and expanded institutional coverage for cross-market studies.

\section*{Acknowledgments}
This work was supported by the National Natural Science Foundation of China (Grant 62272210).

\bibliographystyle{elsarticle-num} 
\bibliography{reference}

\end{document}